\documentclass[
    reprint,
    superscriptaddress,
    aps, prl,
    amsmath,amssymb,
    twocolumn,
    showkeys,
    floatfix,
    nofootinbib,
]{revtex4-2}

\newcommand{\papertitle}{Challenging Spontaneous Quantum Collapse with XENONnT}
\newcommand{\ca}{{\sim}}

\usepackage{graphicx}% Include figure files
\usepackage{dcolumn}% Align table columns on decimal point
\usepackage{bm}% bold math
\usepackage[mathlines]{lineno}% Enable numbering of text and display math
\usepackage{siunitx}
\usepackage{tabularx}
\usepackage{booktabs}
\usepackage{orcidlink} % For ORCID links in author list

\usepackage[]{hyperref}% add hypertext capabilities
\hypersetup{
    colorlinks=true,
    linkcolor=blue,
    filecolor=blue,      
    urlcolor=blue,
    citecolor=blue,
    pdftitle={\papertitle}
    }
\begin{document}

%\preprint{APS/123-QED}

\title{\papertitle}% Force line breaks with \\

\newcommand{\bologna}{\affiliation{Department of Physics and Astronomy, University of Bologna and INFN-Bologna, 40126 Bologna, Italy}}
\newcommand{\chicago}{\affiliation{Department of Physics, Enrico Fermi Institute \& Kavli Institute for Cosmological Physics, University of Chicago, Chicago, IL 60637, USA}}
\newcommand{\coimbra}{\affiliation{LIBPhys, Department of Physics, University of Coimbra, 3004-516 Coimbra, Portugal}}
\newcommand{\columbia}{\affiliation{Physics Department, Columbia University, New York, NY 10027, USA}}
\newcommand{\lngs}{\affiliation{INFN-Laboratori Nazionali del Gran Sasso and Gran Sasso Science Institute, 67100 L'Aquila, Italy}}
\newcommand{\mainz}{\affiliation{Institut f\"ur Physik \& Exzellenzcluster PRISMA$^{+}$, Johannes Gutenberg-Universit\"at Mainz, 55099 Mainz, Germany}}
\newcommand{\mpik}{\affiliation{Max-Planck-Institut f\"ur Kernphysik, 69117 Heidelberg, Germany}}
\newcommand{\munster}{\affiliation{Institut f\"ur Kernphysik, University of M\"unster, 48149 M\"unster, Germany}}
\newcommand{\nikhef}{\affiliation{Nikhef and the University of Amsterdam, Science Park, 1098XG Amsterdam, Netherlands}}
\newcommand{\nyuad}{\affiliation{New York University Abu Dhabi - Center for Astro, Particle and Planetary Physics, Abu Dhabi, United Arab Emirates}}
\newcommand{\purdue}{\affiliation{Department of Physics and Astronomy, Purdue University, West Lafayette, IN 47907, USA}}
\newcommand{\rice}{\affiliation{Department of Physics and Astronomy, Rice University, Houston, TX 77005, USA}}
\newcommand{\stockholm}{\affiliation{Oskar Klein Centre, Department of Physics, Stockholm University, AlbaNova, Stockholm SE-10691, Sweden}}
\newcommand{\subatech}{\affiliation{SUBATECH, IMT Atlantique, CNRS/IN2P3, Nantes Universit\'e, Nantes 44307, France}}
\newcommand{\torino}{\affiliation{INAF-Astrophysical Observatory of Torino, Department of Physics, University  of  Torino and  INFN-Torino,  10125  Torino,  Italy}}
\newcommand{\ucsd}{\affiliation{Department of Physics, University of California San Diego, La Jolla, CA 92093, USA}}
\newcommand{\wis}{\affiliation{Department of Particle Physics and Astrophysics, Weizmann Institute of Science, Rehovot 7610001, Israel}}
\newcommand{\zurich}{\affiliation{Physik-Institut, University of Z\"urich, 8057  Z\"urich, Switzerland}}
\newcommand{\paris}{\affiliation{LPNHE, Sorbonne Universit\'{e}, CNRS/IN2P3, 75005 Paris, France}}
\newcommand{\freiburg}{\affiliation{Physikalisches Institut, Universit\"at Freiburg, 79104 Freiburg, Germany}}
\newcommand{\napels}{\affiliation{Department of Physics ``Ettore Pancini'', University of Napoli and INFN-Napoli, 80126 Napoli, Italy}}
\newcommand{\nagoya}{\affiliation{Kobayashi-Maskawa Institute for the Origin of Particles and the Universe, and Institute for Space-Earth Environmental Research, Nagoya University, Furo-cho, Chikusa-ku, Nagoya, Aichi 464-8602, Japan}}
\newcommand{\laquila}{\affiliation{Department of Physics and Chemistry, University of L'Aquila, 67100 L'Aquila, Italy}}
\newcommand{\tokyo}{\affiliation{Kamioka Observatory, Institute for Cosmic Ray Research, and Kavli Institute for the Physics and Mathematics of the Universe (WPI), University of Tokyo, Higashi-Mozumi, Kamioka, Hida, Gifu 506-1205, Japan}}
\newcommand{\kobe}{\affiliation{Department of Physics, Kobe University, Kobe, Hyogo 657-8501, Japan}}
\newcommand{\kit}{\affiliation{Institute for Astroparticle Physics, Karlsruhe Institute of Technology, 76021 Karlsruhe, Germany}}
\newcommand{\tsinghua}{\affiliation{Department of Physics \& Center for High Energy Physics, Tsinghua University, Beijing 100084, P.R. China}}
\newcommand{\ferrara}{\affiliation{INFN-Ferrara and Dip. di Fisica e Scienze della Terra, Universit\`a di Ferrara, 44122 Ferrara, Italy}}
\newcommand{\groningen}{\affiliation{Nikhef and the University of Groningen, Van Swinderen Institute, 9747AG Groningen, Netherlands}}
\newcommand{\westlake}{\affiliation{Department of Physics, School of Science, Westlake University, Hangzhou 310030, P.R. China}}
\newcommand{\shenzhen}{\affiliation{School of Science and Engineering, The Chinese University of Hong Kong (Shenzhen), Shenzhen, Guangdong, 518172, P.R. China}}
\newcommand{\coimbrapoli}{\affiliation{Coimbra Polytechnic - ISEC, 3030-199 Coimbra, Portugal}}
\newcommand{\uniheidelberg}{\affiliation{Physikalisches Institut, Universit\"at Heidelberg, Heidelberg, Germany}}
\newcommand{\roma}{\affiliation{INFN-Roma Tre, 00146 Roma, Italy}}
\newcommand{\bucknell}{\affiliation{Department of Physics \& Astronomy, Bucknell University, Lewisburg, PA, USA}}

% External affiliations
\newcommand{\fermi}{\affiliation{Centro Ricerche Enrico Fermi—Museo Storico della Fisica e Centro Studi e Ricerche ``Enrico Fermi'', Rome, Italy}}
\newcommand{\frascati}{\affiliation{Laboratori Nazionali di Frascati, INFN, Frascati, Italy}}
\newcommand{\ifin}{\affiliation{IFIN-HH, Institutul National pentru Fizica si Inginerie Nucleara Horia Hulubei, M\v{a}gurele, Romania}}
% End of AFFILIATIONS
\author{E.~Aprile\,\orcidlink{0000-0001-6595-7098}}\columbia
\author{J.~Aalbers\,\orcidlink{0000-0003-0030-0030}}\groningen
\author{K.~Abe\,\orcidlink{0009-0000-9620-788X}}\tokyo
\author{S.~Ahmed Maouloud\,\orcidlink{0000-0002-0844-4576}}\paris
\author{L.~Althueser\,\orcidlink{0000-0002-5468-4298}}\munster
\author{B.~Andrieu\,\orcidlink{0009-0002-6485-4163}}\paris
\author{E.~Angelino\,\orcidlink{0000-0002-6695-4355}}\torino\lngs
\author{D.~Ant\'on~Martin\,\orcidlink{0000-0001-7725-5552}}\chicago
\author{S.~R.~Armbruster\,\orcidlink{0009-0009-6440-1210}}\mpik
\author{F.~Arneodo\,\orcidlink{0000-0002-1061-0510}}\nyuad
% ...
\collaboration{XENON Collaboration}\email[]{xenon@lngs.infn.it}\noaffiliation

\author{C.~Curceanu\,\orcidlink{0000-0002-1990-0127}}\frascati\ifin
\author{S.~Manti\,\orcidlink{0000-0003-3770-0863}}\frascati
\author{K.~Piscicchia\,\orcidlink{0000-0001-6879-452X}}\email[]{kristian.piscicchia@cref.it}\fermi\frascati
\date{\today}% It is always \today, today,

\begin{abstract}
We report on the search for \mbox{X-ray} radiation as predicted from dynamical quantum collapse with low-energy electronic recoil data in the energy range of \SIrange[]{1}{140}{keV} from the first science run of the XENONnT dark matter detector. Spontaneous radiation is an unavoidable effect of dynamical collapse models, which were introduced as a possible solution to the long-standing measurement problem in quantum mechanics. The analysis utilizes a model that for the first time accounts for cancellation effects in the emitted spectrum, which arise in the X-ray range due to the opposing electron-proton charges in xenon atoms. New world-leading limits on the free parameters of the Markovian continuous spontaneous localization and Diósi-Penrose models are set, improving previous best constraints by two orders of magnitude and a factor of five, respectively. 
For the strength and correlation length of the continuous spontaneous localization model, values in the originally proposed parameter ranges are experimentally excluded for the first time.
\end{abstract}

\maketitle

%\tableofcontents

%{\it Introduction.} --- 
One of the greatest mysteries of quantum mechanics is how classical behavior emerges from the quantum realm. The superposition principle has been experimentally verified at the microscopic level with extreme precision~\cite{Fein:2019dgf}, but superpositions of macroscopic states are not observed. Moreover, the mechanism that triggers the quantum-to-classical transition is not encoded in quantum mechanics. This \emph{measurement problem} has led to the development of consistent phenomenological theories that account for the progressive breakdown of quantum superpositions as the size and complexity of a system increase. Dynamical Collapse Models (DCMs) consist in non-linear and stochastic modifications of the Schr\"{o}dinger equation, which require the introduction of phenomenological parameters, in such a way that the collapse rate results to be extremely small for microscopic objects, but an amplification mechanism ensures localization at the macroscopic level.  The modified dynamics result in a Brownian-like diffusion of quantum systems in space during which charged particles emit \emph{spontaneous electromagnetic radiation}. DCMs have been tested using interferometry~\cite{Bassi:2012bg,Carlesso:2022pqr}, by measuring the heating of astrophysical objects~\cite{Ocampo:2024dhj} and by trying to detect indirect effects of collapse dynamics using systems such as cold atoms~\cite{Bilardello:2016bou}, optomechanical devices~\cite{Vinante:2020gow, Pontin:2019bpg}, phonon excitations in crystals~\cite{Adler:2018bul}, and gravitational wave detectors~\cite{Carlesso:2016khv, Helou:2016rii, Carlesso:2017bzy, Armano:2018kix}.  Moreover, searches for spontaneous radiation in the form of \mbox{X-rays} or \mbox{$\gamma$-rays} have been carried out with germanium detectors~\cite{Majorana:2022xuq,Donadi:2020kzc,Donadi:2021myk}.

In this Letter, we report on the search for spontaneous radiation emission by xenon atoms with the XENONnT dark matter detector, as predicted by two benchmark DCM models: the continuous spontaneous localization (CSL) \cite{Pearle:1988uh,Ghirardi:1989cn} and Di\'osi-Penrose (DP) \cite{Diosi:1986nu,Diosi:1989hlx,Penrose:1996cv} models. The two models share a similar structure for the dynamical equation. In the Markovian formulation, which is investigated in this work, a white stochastic noise is assumed in time. The models are characterized by a collapse rate parameter which sets the strength of the collapse, and a correlation length which measures the spatial resolution of
the collapse, setting the scale beyond which spatial superpositions are suppressed. The key distinction between the two models lies in their spatial correlation forms: for CSL, it is Gaussian, with an amplitude determined by the collapse strength $\lambda$ and a resolution scale set by $r_{\mathrm{C}}$. Ghirardi, Rimini, and Weber (GRW) \cite{Ghirardi:1985mt} proposed values on the order of magnitude $\lambda^{\mathrm{GRW}} \simeq \SI{e-16}{s^{-1}}$ and $r_{\mathrm{C}}^{\mathrm{GRW}} \simeq \SI{e-7}{m}$, which ensure the effective collapse of macroscopic systems. 

The collapse in the DP model is linked to gravity, and the spatial correlation is proportional to the Newtonian potential. The correlation length $R_0$ acts as a spatial cutoff parameter to avoid a divergence in the collapse rate for point-like particles, while the strength is given by the gravitational constant $G$. Theoretical upper bounds on $R_0$ can be set by requiring the prevention of macroscopic quantum superpositions~\cite{Figurato:2024tpo} before they can be perceived. This sets the collapse timescale governed by $R_0$ at $\lesssim\SI{0.01}{s}$, the approximate time resolution of the human eye. For the model systems investigated in~\cite{Figurato:2024tpo}, the limits depend strongly on their size $L$: $R_0 \lesssim 10^3\;\text{\AA}$ for $L \sim \SI{1}{\micro m}$ and $R_0 \lesssim 5 \times 10^6\;\text{\AA}$ for $L \sim \SI{10}{\micro m}$.

Spontaneous radiation searches provide the strongest experimental bounds on the model parameters $R_0$ of DP, as well as on the $\lambda$ and $r_{\mathrm{C}}$ of CSL over a broad range of the parameter space~\cite{Donadi:2020kzc, Donadi:2021myk,Majorana:2022xuq}. At sufficiently high energies of $\mathcal{O}(100\,\text{keV}-\text{MeV})$, the spontaneous radiation wavelengths $\lambda_\gamma$ are intermediate between the nuclear and atomic length scales. With good approximation, protons would emit radiation coherently, while electrons would emit incoherently~\cite{Donadi:2021myk}, leading to
\begin{equation}\label{eq:rate_high_energy}
\left. \frac{d\Gamma}{dE} \right|_t^{\mathrm{CSL,DP}} \propto E^{-1} \left( N_{\mathrm{p}}^2 + N_{\mathrm{e}}  \right),
\end{equation}
for the rate of the spontaneous radiation. Here, $N_{\mathrm{p}}$ and $N_{\mathrm{e}}$ are the numbers of protons and electrons in the atom while $E$ is the photon energy for the radiation emitted. The constant of proportionality depends on the respective model parameters. Strong constraints on collapse models in the \mbox{X-ray} regime (\mbox{\SIrange[]{19}{100}{keV}}) were previously set in Ref.~\cite{Majorana:2022xuq}, where the spontaneous emission from germanium atoms of the target was modeled according to the high-energy approximation as per Eq.~\eqref{eq:rate_high_energy}. However, the phenomenology of spontaneous radiation is more complex when \mbox{X-rays} are considered, with $\lambda_\gamma$ of the order of the atomic orbit dimensions. Here, coherent emission by the electrons and cancellation effects between the oppositely charged electrons and protons must be considered. The generalization of the emission rates for energies as low as \SI{1}{keV} was derived in Ref.~\cite{Piscicchia:2024wan}. We use these results to search for spontaneous radiation emission signals in the energy range of \mbox{\SIrange[]{1}{140}{keV}}. The implementation of the theoretical framework for this analysis is described in the supplemental material~\cite{SM}.

%{\it The XENONnT experiment.} ---
Due to their low background, low energy threshold of \SI{\ca 1}{keV} \cite{XENON:2022ltv}, and target mass of several tonnes, contemporary dual-phase xenon time projection chambers (TPCs) promise high sensitivities to spontaneous radiation, by using liquid xenon (LXe) both as emitter and detection medium. In these detectors, energy deposited from interactions with the xenon target atoms gives rise to prompt scintillation light and free ionization electrons. The former is directly detected as so-called S1 signal by arrays of photomultiplier tubes (PMTs) at the top and bottom of the TPC. The liberated charge is converted into a likewise detectable secondary scintillation signal (S2). For this, an electric field between a cathode and gate electrode drifts the electrons toward the top of the TPC. There, a second, stronger electric field between the gate and an anode extracts and accelerates them into the gaseous phase, resulting in electroluminescence. Pairing both signals enables the reconstruction of the deposited energy and the interaction position. 
The combined energy scale $E$ is computed from the corrected scintillation signals ($cS1$, $cS2$) as $E = W(cS1/g_1 + cS2/g_2)$, where the mean energy to produce a quantum $W = \SI{13.7}{eV/quanta}$ \cite{Dahl:2009nta}, and the detector-specific gain constants $g_1$ and $g_2$ are considered \cite{XENON:2024qgt}.
In general, the charge-to-light ratio is larger for particle interactions with the electrons as opposed to the nucleus of the xenon atom. This enables discrimination between electronic recoil (ER) and nuclear recoil (NR) interactions \cite{Aprile:2006kx}, with the former being expected for the \mbox{X-ray} signatures investigated in this Letter. 

The XENONnT experiment \cite{XENON:2024wpa} features a  dual-phase TPC with a sensitive LXe mass of \SI{5.9}{t}, monitored by a total of 494 Hamamatsu R11410-21 PMTs \cite{Antochi:2021wik}. It offers an unprecedentedly low ER background of \SI{15.8(13)}{events/(t.y.keV)} below recoil energies of \SI{30}{keV} in its first science run \cite{XENON:2022ltv}. To suppress cosmogenic backgrounds, this experiment is located at the INFN Laboratori Nazionali del Gran Sasso (LNGS) in Italy, with a rock overburden providing shielding equivalent to about \SI{3600}{meters} of water. Further background reduction is achieved through active tagging using a water tank surrounding the TPC, which is instrumented as a muon veto~\cite{XENON1T:2014eqx} and supplemented by an inner neutron veto system~\cite{XENON:2024fxf}. Moreover, an extensive radioassay campaign for the installed detector materials was performed \cite{XENON:2021mrg} and the xenon in the detector is continuously purified from chemical and radioactive impurities \cite{Plante:2022khm, XENON:2016bmq, Murra:2022mlr}.

%{\it Spontaneous radiation searches with XENONnT.} ---
With a few exceptions detailed below, the search for the spontaneous radiation signatures in ER data from XENONnT follows the analysis performed in Ref.~\cite{XENON:2022ltv}. The signal reconstruction, calibration, and event selection are further detailed in Ref.~\cite{XENON:2024qgt}. Data from the first XENONnT science run (SR0), collected from July 6, 2021, to November 10, 2021, with an exposure of \SI{1.16}{t.y} inside the \SI{4.37(14)}{t} fiducial mass, is employed. 

The expected signal spectra are modeled from the theoretical shapes according to Ref.~\cite{Piscicchia:2024wan}. Details on the signal models can be found in the supplemental material~\cite{SM}. The detector response is modeled by a skew-Gaussian energy resolution smearing and the application of the combined efficiency of detection and event selection~\cite{XENON:2022ltv} to the signal spectra. Compared to ~\cite{XENON:2022ltv}, there are two differences in the efficiency modeling: First, work considers an acceptance loss from a data-selection against accidental pairings of isolated light and charge signals that was underestimated previously.
Second, the event-building efficiency, determined analogously to Ref.~\cite{XENON:2025vwd} and projected to energy space, is considered. These efficiency adjustments influence the CSL [DP] model results in the following by \SI{\ca 17}{\percent} [\SI{\ca 6}{\percent}].

A total of nine components are considered in the background model. These encompass the 
double-beta decay of $^{136}$Xe, the beta decays of $^{214}$Pb, $^{133}$Xe, and $^{85}$Kr, the double-electron capture of $^{124}$Xe, and the decay of $^{83\mathrm{m}}$Kr via internal conversion inside the LXe target.
Other backgrounds are gamma rays from detector construction materials, solar neutrinos scattering off electrons, and random pairings of uncorrelated S1 and S2s, referred to as accidental coincidences (ACs). 
Analogously to Ref.~\cite{XENONCollaboration:2023dar}, the free electron approximation in the solar neutrino spectrum used in Ref.~\cite{XENON:2022ltv} is supplemented by a stepping approximation \cite{Chen:2016eab}, accounting for the electron binding energies. Furthermore, the relative $^{85}$Kr rate uncertainty, which was previously overestimated due to an erroneous unit in the uncertainty propagation, was reduced from \SI{\ca 65}{\percent} to \SI{\ca 22}{\percent}. The impact of the adjustments to the background model compared to Ref.~\cite{XENON:2022ltv} on the results obtained in the following is \SI{< 1}{\percent} for the CSL and DP models.

\begin{figure}
    \centering
    \includegraphics[width=1\linewidth]{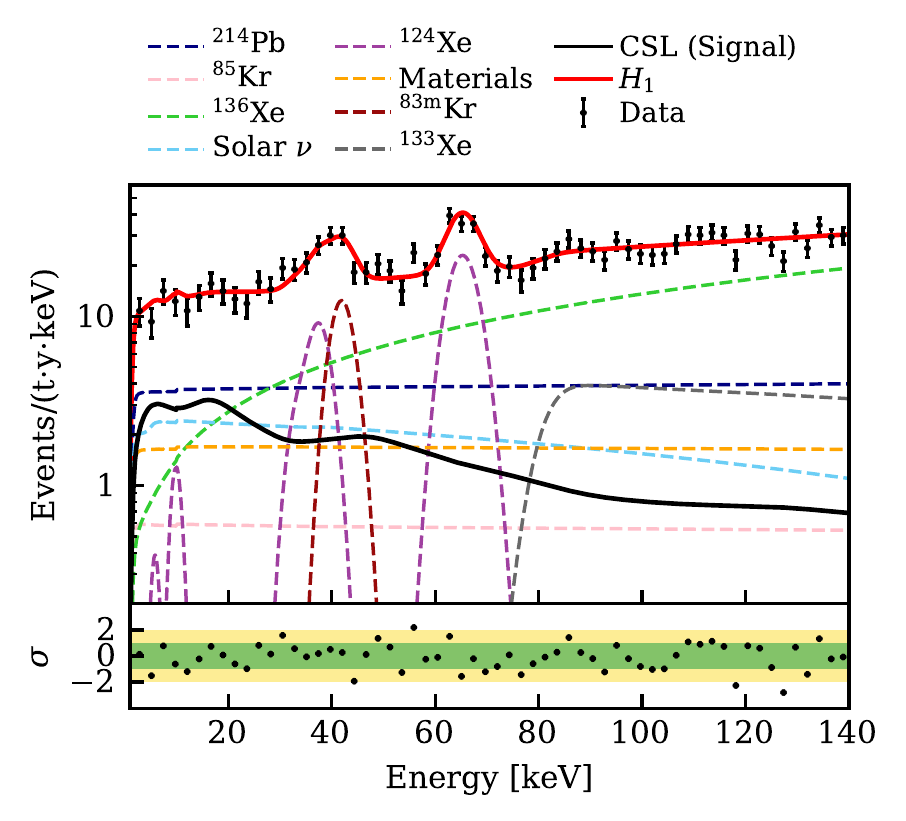}
    \caption{Best-fit spectrum for the combined CSL signal and background model ($H_1$) to the XENONnT SR0 low-energy electronic recoil data. The individual background components are indicated by dashed lines, except for the subdominant AC background spectrum which is out of range and, therefore, not shown. The best-fit result for the DP model is practically identical (see also \autoref{fig:wfc_csl_dp_spectra}). The model discontinuity around 10~keV is due to the blinding of the dark matter search region %in the S1-S2 parameter space 
    for~\cite{XENON:2023cxc} which marginally reduces the ER rate in the data from~\cite{XENON:2022ltv} at the lowest energies.}
    \label{fig:csl_fit_spectra}
\end{figure}
The fit of the combined signal and background model ($H_1$) to the measured data is performed in the parameter space of reconstructed energy in the range of \mbox{\SIrange[]{1}{140}{keV}} employing an unbinned maximum likelihood analogous to Refs.~\cite{XENON:2020rca, XENON:2022ltv}. The best-fit result for the Markovian CSL model is shown in \autoref{fig:csl_fit_spectra}. The low-energy part of the signal region contains the highest rates and distinguishing features of the spontaneous emission spectra (see supplemental material~\cite{SM}). This allows us to place stringent limits on the respective model parameters using a log-likelihood ratio test statistic and assuming asymptoticity of the likelihood \cite{Baker:1983tu}.

For the Markovian CSL model, $\lambda/r_{\mathrm{C}}^2$ does not change the spectral shape, so it can be deduced directly from the obtained fit rate multiplier when assuming a fixed initial value for $\lambda/r_{\mathrm{C}}^2$. The CSL best-fit result is shown in \autoref{fig:csl_fit_spectra}, with a signal rate of $50^{+200}_{-50}\;\text{events}/(\text{t}\times\text{y})$ and a \mbox{p-value} of $\sim 0.3$. 
As a local discovery significance of $0.3\,\sigma$ is obtained, no significant excess is found and an upper limit of $\lambda/r_{\mathrm{C}}^2 < \SI{3.0e-3}{s^{-1}m^{-2}}$ [$\SI{3.7e-3}{s^{-1}m^{-2}}$] at \SI{90}{\percent} confidence level (C.L.) [\SI{95}{\percent} C.L.] is derived. This constitutes an improvement over the previous best limit from the \textsc{Majorana Demonstrator} \cite{Majorana:2022xuq} by a factor of $\ca 135$ and sets the new most stringent experimental constraint for $r_{\mathrm{C}} \lesssim \SI{e-5}{m}$. A comparison of these limits with constraints from other experiments and theoretical considerations is given in \autoref{fig:wfc_csl_limits_95cl}.  
According to the log-likelihood ratio test statistic, the parameter set ($r_{\mathrm{C}}^{\mathrm{GRW}}=\SI{1e-7}{m},\;\lambda^{\mathrm{GRW}}$=\SI{1e-16}{s^{-1}}) deviates from the best-fit result by $9.1\,\sigma$. For the first time, this allows to experimentally exclude CSL model parameter values in the range originally proposed by GRW. 
\begin{figure}
    \centering
    \includegraphics[width=1\linewidth]{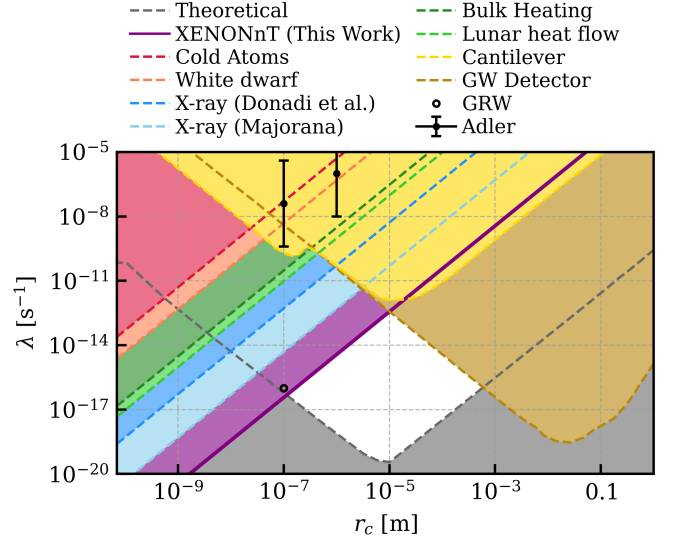}
    \caption{XENONnT \SI{95}{\percent} C.L.\  upper bound on the CSL model parameters, compared to exclusion limits from other non-interferometric experiments and theoretical propositions, with the shaded areas marking the excluded regions of the parameter space. Results from the \textsc{LISA Pathfinder} gravitational wave detector \cite{Carlesso:2017bzy, Armano:2018kix}, microcantilever measurements \cite{Vinante:2020gow}, cold atoms experiments \cite{Bilardello:2016bou}, and estimates from the bulk heating rate in the \textsc{CUORE} experiment \cite{CUORE:2017ztm} are given. The two previous best limits from \mbox{X-ray} emission searches by Donadi et al.\ \cite{Donadi:2021myk} and the \textsc{Majorana Demonstrator} \cite{Majorana:2022xuq} are shown as well. Moreover, astrophysical bounds from lunar thermal emission as well as from the cooling of white dwarf J1251+4403 are indicated~\cite{Ocampo:2024dhj}. A commonly adopted lower theoretical bound can be introduced by requiring a maximum collapse time of \SI{\ca 10}{ms}, i.e., the perception time of the human eye, for a superposition of a single-layered graphene disk of a barely visible radius of \SI{\ca 10}{\micro m} \cite{Toros:2017col}. The black markers indicate theoretical suggestions by Adler \cite{Adler:2006nh} and Ghirardi, Rimini, and Weber \cite{Ghirardi:1986uni}.}
    \label{fig:wfc_csl_limits_95cl}
\end{figure}

Due to the more complex rate dependencies on the free parameter $R_0$ in the DP model, which can result in changes of the spectral shape beyond a global scaling, the fitting method employing a single linear rate multiplier requires an iterative approach for the DP model: A fixed starting value for $R_0$ and, thus, a static spectral shape are assumed. The values assumed in the next iterations are then obtained from the best-fit rate multiplier of the previous iteration by considering only a scaling by the leading term proportional to $R_0^{-3}$ (see Ref.~\cite{Piscicchia:2024wan}). 
Any starting values that are not excluded by former experiments lead to the convergence to a best-fit value of $R_0 \approx \SI{9.1e-10}{m}$. Around this value, the spectral shape of the signal does not change significantly (see Figure~S-1 in supplemental material~\cite{SM}), such that best-fit values, significances, and limits are in good approximation independent of the assumed $R_0$. 
With a local DP discovery significance of $0.2\,\sigma$, a lower limit of $R_0 > \SI{4.9e-10}{m}$ [$\SI{4.5e-10}{m}$] at \SI{90}{\percent} C.L.\ [\SI{95}{\percent} C.L.] is set. This constitutes an improvement by approximately a factor of five compared to the previous world-leading limit by the \textsc{Majorana Demonstrator} \cite{Majorana:2022xuq}.

%{\it Summary and Conclusion.} ---
In summary, new best constraints on the CSL and DP model, respectively, of $\lambda/r_{\mathrm{C}}^2 < \SI{3.0e-3}{s^{-1}m^{-2}}$ [$\SI{3.7e-3}{s^{-1}m^{-2}}$] and $R_0 > \SI{4.9e-10}{m}$ [$\SI{4.5e-10}{m}$] at \SI{90}{\percent} C.L.\ [\SI{95}{\percent} C.L.] are set. The former allows for the first-time exclusion of the commonly assumed values ($r_{\mathrm{C}}^{\mathrm{GRW}}, \lambda^{\mathrm{GRW}}$) for a white noise field. 
The limit enhancements demonstrate the broad physics reach of the XENONnT experiment and dual-phase xenon TPCs in general. They further drive the ongoing exploration of the still feasible parameter space for the most well-studied dynamical collapse models. 
A wider parameter range remains to be probed for colored or dissipative model extensions \cite{Adler:2007col, Toros:2017col}. 
With more XENONnT data already available, further sensitivity enhancements are expected. 

We gratefully acknowledge support from the National Science Foundation, Swiss National Science Foundation, German Ministry for Education and Research, Max Planck Gesellschaft, Deutsche Forschungsgemeinschaft, Helmholtz Association, Dutch Research Council (NWO), Fundacao para a Ciencia e Tecnologia, Weizmann Institute of Science, Binational Science Foundation, Région des Pays de la Loire, Knut and Alice Wallenberg Foundation, Kavli Foundation, JSPS Kakenhi, JST FOREST Program, and ERAN in Japan, Tsinghua University Initiative Scientific Research Program, UZH Postdoc Grant (Grant No. FK-24-101), DIM-ACAV+ Région Ile-de-France, and Istituto Nazionale di Fisica Nucleare. This project has received funding/support from the European Union’s Horizon 2020 research and innovation program under the Marie Skłodowska-Curie grant agreement No 860881-HIDDeN. C.\ C., K.\ P.\ and S.\ M.\ acknowledge of the John Templeton Foundation, Grant 62099. The opinions expressed in this publication are those of the authors and do not necessarily reflect the views of the John Templeton Foundation. C.\ C.\  acknowledge support from the Foundational Questions Institute and Fetzer Franklin Fund, a donor advised fund of Silicon Valley Community Foundation (Grants No. FQXi-RFP-CPW-2008 and FQXi-MGB-2011), and from the INFN (VIP).  C.\ C.\ benefited from the EU COST Actions CA23115 and CA23130. K.\ P.\ acknowledge support from the Centro Ricerche Enrico Fermi - Museo Storico della Fisica e Centro Studi e Ricerche “Enrico Fermi” (Open Problems in Quantum Mechanics project). 

We gratefully acknowledge support for providing computing and data-processing resources of the Open Science Pool and the European Grid Initiative, at the following computing centers: the CNRS/IN2P3 (Lyon - France), the Dutch national e-infrastructure with the support of SURF Cooperative, the Nikhef Data-Processing Facility (Amsterdam - Netherlands), tfhe INFN-CNAF (Bologna - Italy), the San Diego Supercomputer Center (San Diego - USA) and the Enrico Fermi Institute (Chicago - USA). We acknowledge the support of the Research Computing Center (RCC) at The University of Chicago for providing computing resources for data analysis.

We thank the INFN Laboratori Nazionali del Gran Sasso for hosting and supporting the XENON project.
%\appendix
%
%\section{Appendices}
%
%...

%\nocite{*}
\nocite{coulson, gill2003, Kohn:1965zzb, Mortensen:2024gpa}

\bibliography{references}% Produces the bibliography via BibTeX.

\clearpage
\appendix
\section{Supplemental Material}
\setcounter{equation}{0}
\setcounter{table}{1}
\setcounter{page}{1}
\makeatletter
\renewcommand{\theequation}{S\arabic{equation}}
\renewcommand{\thefigure}{S\the\numexpr\value{figure}-4\relax} % we have to manually set this because the hyperlink will fail if we simply \setcounter{figure}{0}
\renewcommand{\thetable}{S\the\numexpr\value{table}-1\relax}

The low-energy generalization of the spontaneous radiation emission rates for the CSL and DP models used in this work was derived in Ref.~\cite{Piscicchia:2024wan}. The expression for the expected DP emission rate can be found in Ref.~\cite{Piscicchia:2024wan}. It follows the Diósi \cite{Diosi:1986nu, Diosi:1989hlx} as opposed to the  Penrose \cite{Penrose:1996cv} convention, which differ by a scaling factor of $8\pi$.

The expression for the Markovian CSL emission rate was simplified to
\begin{widetext}
\begin{align}\label{eq:rate_csl_detailed_limit}
	\begin{split}
		\left. \dfrac{\mathrm{d}\Gamma}{\mathrm{d}E} \right\rvert_t^{\mathrm{CSL}} = &\ \dfrac{\hbar e^2 \lambda}{4 \pi^2 \varepsilon_0 c^3 m_0^2 r_{\mathrm{C}}^2 E} \cdot \left\{\vphantom{\dfrac{E}{h}} \right.
		N_{\mathrm{p}}^2 + N_{\mathrm{e}}  + 2 \cdot \sum_{o\,o'\,\mathrm{pairs}} N_o N_{o'} \cdot {\mathrm{sinc}{\left(\frac{\beta |\rho_o - \rho_{o'}|E }{\hbar c}\right)}}\\
		&- 2 N_{\mathrm{p}} \cdot \sum_{o} N_o \cdot {\mathrm{sinc}{\left(\frac{\rho_o E}{\hbar c}\right)}} + \sum_{o} N_o (N_o-1) \cdot {\mathrm{sinc}{\left(\frac{\alpha \rho_o E}{\hbar c}\right)}}
		\left . \vphantom{\dfrac{E}{h}}\right\},
	\end{split}
\end{align}
\end{widetext}
with $\textrm{sinc}{(x)} = \sin (x) / x$ for $x>0$.
The simplification follows from the scale parameter $r_{\mathrm{C}}\gtrsim \SI{1.15e-8}{m}$ significantly exceeding the mutual distances between all charges in the atomic system~\cite{Donadi:2021myk, Toros:2017col}. 
In \autoref{eq:rate_csl_detailed_limit} the distances between electron pairs are parameterized in terms of atomic orbit mean radii ($\rho_o$) using the constants $\alpha$ and $\beta$.
The values $\alpha \in [1,1.5]$ and $\beta = 1.04$ are estimated from available calculations of the average electron distances, based on position intracule functions \cite{coulson,gill2003}. $N_o$ represents the number of electrons in the $o$-th orbit of the atom.
The mean radii of the xenon atomic orbits were calculated based on a density functional theory (DFT) \cite{Kohn:1965zzb} all-electron calculation with the DFT code GPAW \cite{Mortensen:2024gpa}. The results are shown in \autoref{tab:xe_orbitals}.

The CSL spectrum expected for ($r_{\mathrm{C}}^{\mathrm{GRW}}, \lambda^{\mathrm{GRW}}$), as well as the best-fit spectrum, are shown in \autoref{fig:wfc_csl_dp_spectra}. The latter is practically identical to the indicated DP best-fit spectrum.
Within the constraint of $\alpha \in [1,1.5]$, the expected integrated signal rate varies by at most \SI{\ca 1}{\percent} compared to the rate at $\alpha = 1.25$ assumed in the analysis. The limits change by \SI{\ca 7}{\percent} for the CSL model and $\ca \SI{2}{\percent}$ in the DP case.

\begin{table}%[h]
     \centering
     \caption{Orbital properties of the xenon atom.}
     \begin{tabular}{cc|c} % {.9\linewidth}
         \toprule
         \textbf{State} & \textbf{Occupation} &
\textbf{Mean Radius $\rho_o$ [$\mathbf{10^{-10}}$\,m]} \\
         \midrule
         1s & $\phantom{1}2$ & $0.015$ \\
         2s & $\phantom{1}2$ & $0.064$ \\
         2p & $\phantom{1}6$ & $0.055$ \\
         3s & $\phantom{1}2$ & $0.170$ \\
         3p & $\phantom{1}6$ & $0.165$ \\
         3d & $10$ & $0.149$ \\
         4s & $\phantom{1}2$ & $0.395$ \\
         4p & $\phantom{1}6$ & $0.412$ \\
         4d & $10$ & $0.461$ \\
         5s & $\phantom{1}2$ & $1.024$ \\
         5p & $\phantom{1}6$ & $1.231$ \\
         \bottomrule
     \end{tabular}
     \label{tab:xe_orbitals}
\end{table}
\begin{figure}
    \centering
    \includegraphics[width=1\linewidth]{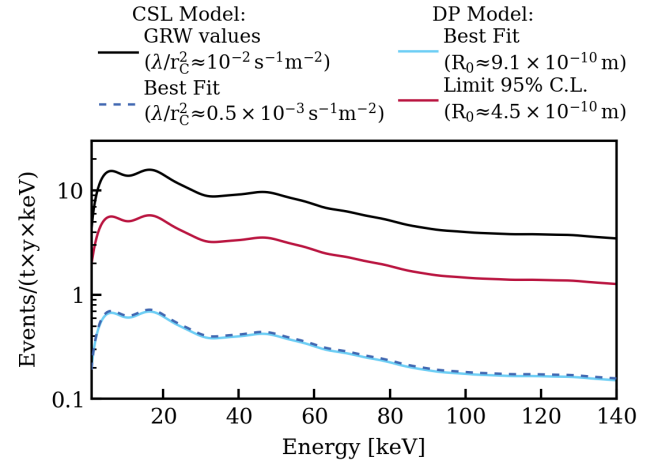}
    \caption{Spontaneous radiation rate spectrum (black) from xenon atoms for a Markovian CSL model according to \autoref{eq:rate_csl_detailed_limit}, assuming the model parameter values proposed by Ghirardi, Rimini, and Weber \cite{Ghirardi:1986uni}. 
    The CSL best-fit spectrum is indicated as dark blue dashed line. 
    The Markovian DP model spectra from Ref.~\cite{Piscicchia:2024wan} with the best-fit value and \SI{95}{\percent} C.L.\ upper limit of $R_0$ are shown in light blue and red, respectively.}
    \label{fig:wfc_csl_dp_spectra}
\end{figure}

\end{document}
%
% ****** End of file ******